\documentclass[useAMS,fleqn,usenatbib]{mn2e}
\usepackage{amssymb,amsmath}
\usepackage{graphics}

\def\eqref#1{equation~(\ref{#1})}

\def\binom#1#2{{#1 \choose #2}}

\title[Light curve modelling for mutual transits]%
{Light curve modelling for mutual transits}
\author[A. P\'al]{
\noindent Andr\'as P\'al\thanks{E-mail: apal@szofi.net} \\
Konkoly Observatory of the Hungarian Academy of Sciences, 
        Konkoly Thege Mikl\'os \'ut 15-17,
        H-1121 Budapest, Hungary \\
Department of Astronomy, Lor\'and E\"otv\"os University,
        P\'azm\'any P. st. 1/A,
        Budapest H-1117, Hungary}

\begin{document}

\date{Accepted \dots. Received \dots; in original form \dots}

\pagerange{\pageref{firstpage}--\pageref{lastpage}} \pubyear{2011}

\maketitle

\label{firstpage}

\begin{abstract}
In this paper we describe an algorithm and deduce the related
mathematical formulae that allows the computation of observed fluxes 
in stellar and planetary systems with arbitrary number of
bodies being part of a transit or occultation event. The presented
method does not have any limits or constraints for the geometry and
can be applied for almost all of the available limb darkening models as well.
As a demonstration, we apply this scheme to gather information 
for the orbital inclinations
from multiple transiting planetary systems in cases when mutual 
transits occur. We also show here that these mutual events constrain
inclinations unambiguously, yielding a complete picture for the whole system. 
\end{abstract}



\section{Introduction}
\label{sec:introduction}

With the advent of the space missions Corot and Kepler 
\citep[see][]{barge2008,borucki2009} and 
as the result of numerous successful ground-based surveys, 
nearly two-hundred transiting extrasolar planets are known up to date.
Furthermore, the number of candidates awaiting for confirming the
planetary properties exceeds the magnitude of thousand \citep{borucki2011}. 
These systems gives us a unique perspective for various studies because most of 
the planetary and orbital parameters can be obtained without any
ambiguity. Transiting planetary companions are also known in 
multiple stellar systems \citep{doyle2011}. Moreover, both systems of
transiting planets and eclipsing binaries provide substantial information
about the stars from which the absolute physical properties can be 
easily obtained, completely independently from other methods and therefore
these studies are essential to confirm stellar evolution models. 

In the case of multiple transiting planetary 
systems \citep[see e.g.][]{holman2010}, triple or hierarchical
stellar systems or circumbinary planetary systems \citep{doyle2011}, 
planetary systems around one of a binary components or 
systems with planetary companions 
\citep[exomoons, see e.g.][]{szabo2006,simon2009,kipping2009},
there is a chance to observe mutual eclipsing or
transiting events when (at least) three of the bodies are aligned 
along the line of sight. 
Moreover, if planet and/or stellar formation
prefers co-planar orbits, the chance is even higher. Conversely, 
observing mutual transit events yields additional information about the
orbital characteristics of the whole system.
Recently, \cite{sato2009} and
\cite{ragozzine2010} analyzed these effects and their qualitative 
influence on extrasolar moons and multiple planetary systems. 
In this paper we discuss how the photometric measurements are 
affected due to such mutual transiting or eclipsing events by giving an 
algorithm that models light curves of such phenomena. 
Our method described here is capable to compute light curve models for 
arbitrary number of eclipsing or transiting bodies and for all of 
the well-known limb darkening models without any restrictions for 
the projected diameters of the active components.

The structure of this paper is as follows. 
Section~\ref{sec:lightcurvemodel} describes the algorithm and 
the formulae needed to evaluate the fluxes or light curve points
for events with multiple transiting, eclipsing or occulting companions. 
In Section~\ref{sec:mutualtransits} we briefly discuss the qualitative
properties of systems where mutual transits may occur and
demonstrate how information
gathered from such mutual events can be exploited in order to 
constrain orbital alignments in such a transiting planetary system. 
And finally, Section~\ref{sec:summary} summarizes the key points
and results of this paper.


\section{The light curve model}
\label{sec:lightcurvemodel}

In this section we briefly describe the methods used to compute the
light curve models for multiple transiting objects. Recently,
\cite{kipping2011} published an algorithm that is capable
to estimate the observed flux when two bodies transit their host
star simultaneously. However, that method works only when one of these
bodies is very small (i.e. assuming a homogeneous flux density beyond 
this very small disk). Here we demonstrate an alternative algorithm that
is significantly more concise and can be treated as an extension of
the approach by \cite{kipping2011} in several ways. 
First, the presented method
is capable to incorporate more than two transiting bodies.
Space-borne missions like Kepler are expected to detect both extrasolar moons 
\citep{szabo2006,kipping2009} by different methods
(e.g. detecting timing variations or via photometry) and systems
with three or more transiting planetary companions that are also known
\citep[Kepler-11, see][]{lissauer2011}. 
Second, the model can be extended for various limb darkening models,
that can be quantified by a series expansion on the apparent stellar
surface (and might lack circular symmetry). 
Such models can also be exploited to
quantify cases with asymmetric light curves like KOI-13(b) \citep{szabo2011}.
Third, in terms of computation time, the method presented here
can also be an alternative for the well-known models available
for the single-planet cases \citep{mandel2002,gimenez2006,pal2008}.
Also, this method does not require several dozens of 
distinct geometric cases \citep[see][]{mandel2002,kipping2011}.
And finally, more sophisticated cases like non-uniform thermal
radiations can also be considered, even in mutually transiting systems.
This might be relevant in the analysis of near-infrared light curves 
of close-in eclipsing companions. 

The computation of the presented light model is based on 
three subsequent steps. First, a net of disjoint arcs is obtained
from the mutual intersections of the apparent stellar and/or planetary
disc edges. Second, we generate a vector field whose exterior derivative
(i.e. the planar component of the curl operator)
is the surface brightness. The surface brightness must be in accordance
with our assumptions for the limb darkening models.
Third, we apply Green's or the Kelvin-Stokes 
theorem (known from differential geometry or vector calculus)
to integrate the vector field on an appropriately oriented subset 
the arcs. In the following, we discuss these three steps as well as
their applications for various surface brightness functions.

\subsection{Net of arcs}
\label{subsec:arcnet}

In principle, the projected stellar, planetary or lunar discs
are characterized by the center coordinates $x_0,y_0$ and 
the radius $r$. An arc on one of these circles is quantified 
by the additional parameters $\varphi^{(0)}$ and $\Delta \varphi$,
where $\varphi^{(0)}$ is the position angle between the reference
axis ($x+$) and the beginning of the arc while $\Delta \varphi$
is the length of the arc in radians. All of the arcs in this model
are oriented in counter-clockwise (i.e. prograde or positive) direction.
In the following, the circles and arcs are indexed by $k$ and $\ell$,
respectively. Obviously, for the $k$th circle, 
\begin{equation}
\sum\limits_{\ell}\Delta\varphi_{k\ell}=2\pi
\end{equation}
and 
\begin{equation}
\varphi^{(0)}_{k,\ell+1}=\varphi^{(0)}_{k,\ell}+\Delta\varphi_{k,\ell}.
\end{equation}
Completely disjoint circles or circles of which edge does not
intersect other ones have only one arc that is used
to represent the circle itself. Namely, $\{\ell\}=\{1\}$
and $\Delta\varphi_{k,1}=2\pi$ while the value of $\varphi^{(0)}_{k,1}$
can be arbitrary. 

\begin{figure}
\begin{center}
\resizebox{80mm}{!}{\includegraphics{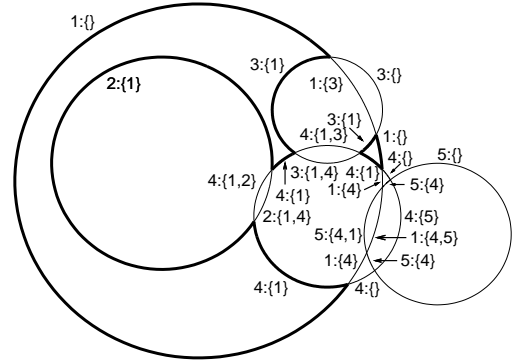}}
\end{center}
\caption{A complex configuration of 5 circles. The
arcs are denoted by $k$:$\{C_{k',\ell}\}$ where $k$ and $k'$ are
the indices of the circles and the set $C_{k',\ell}$ is a 
list of circle indices in which the given arc goes through. The 
thick arcs mark the boundary of the region in circle \#1 that is
disjoint from the other circles. It is easy to see that these
boundary arcs are labelled by 
either $1$:$\{\}$ or $k$:$\{1\}$ (where $2\le k$). }
\label{fig:circles}
\end{figure}

The net of arcs is built iteratively. If the new circle
is disjoint, only one arc is placed with $\Delta\varphi_k,1=2\pi$,
otherwise the two position angles for the two intersection points
are computed using known trigonometric
relations and the appropriate arcs are split into two or three smaller
ones. After obtaining this set of arcs, the topology is also generated.
Namely, by checking for each arc what are the circles that contains this arc
inside. Let us denote this subset of circles regarding to the $\ell$th
arc of the $k$th circle by $C_{k,\ell}$. Note that
this set might be empty or might even contain all of the circle indices
with the exception of $k$. See e.g. Fig.~\ref{fig:circles} for a 
particular example of $5$ intersecting circles. 

\subsection{Surface brightness and vector fields}

The surface brightness of the host star can be modelled by
various limb darkening laws\footnote{See e.g.
http://www.astro.keele.ac.uk/jkt/codes/jktld.html}. 
Recalling Green's theorem over $\mathbb{R}^2$,
we can write 
\begin{equation}
\iint\limits_{S} (\mathbf{D}\wedge \mathbf{f})\,\mathrm{d}A = 
\oint\limits_{\partial S}\mathbf{f}\cdot\mathrm{d}\mathbf{r}.\label{eq:stokes2d}
\end{equation}
Here $S\subset \mathbb{R}^2$ and $\partial S$ is the boundary
of $S$. These are two and one-dimensional manifolds on $\mathbb{R}^2$ 
on which the standard measures are $\mathrm{d}A$ and $\mathrm{d}\mathbf{r}$,
respectively. This equation can also be viewed as a somewhat special case 
to the Stokes theorem known for the curl operator in a three-dimensional space.
The term $\mathbf{D} \wedge \mathbf{f}$ denotes
the exterior derivative of $\mathbf{f}$, that can be written in terms
of vector components as 
\begin{equation}
\mathbf{D}\wedge \mathbf{f} = \frac{\partial f_y}{\partial x}-\frac{\partial f_x}{\partial y}.\label{eq:exterior}
\end{equation}
For our problem discussed in this paper we can apply the above
equation~(\ref{eq:stokes2d}) as follows. First, we have to find a 
function $\mathbf{f}\equiv(f_x,f_y)$ of which exterior derivative is
the given stellar surface brightness density. We have to note
that due to the Young theorem, this function is ambiguous, since
we can add an arbitrary scalar gradient to $\mathbf{f}$,
of which addition does not change its exterior derivative. Therefore,
it is recommended to find such an $\mathbf{f}$ which have ``nice
properties'' making the computation of the integral on the right-hand
side of equation~(\ref{eq:stokes2d}) convenient. For instance,
a homogeneous surface can be modelled with the function
\begin{equation}
\mathbf{f}_{1}=\binom{f_x}{f_y}=\frac{1}{2}\binom{-y}{+x}.\label{eq:unitcurl}
\end{equation}
The exterior derivative of this function is unity and there is no preferred 
direction or position angle in the vector field described by $\mathbf{f}_{1}$.

\subsection{Integration on the arcs}

\begin{figure*}
\begin{center}
\resizebox{20mm}{!}{\includegraphics{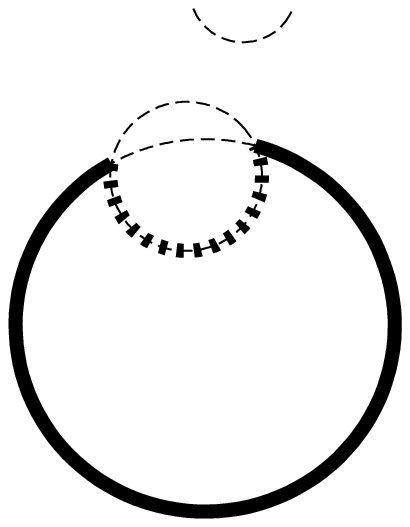}}%
\resizebox{20mm}{!}{\includegraphics{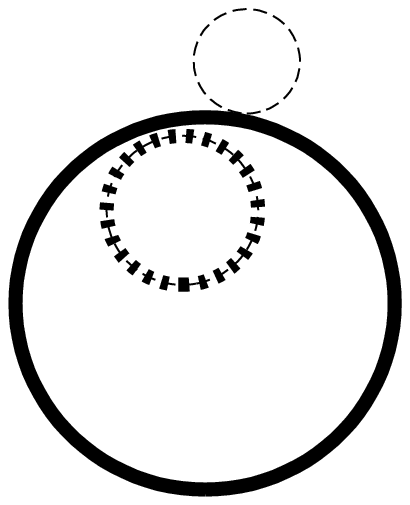}}%
\resizebox{20mm}{!}{\includegraphics{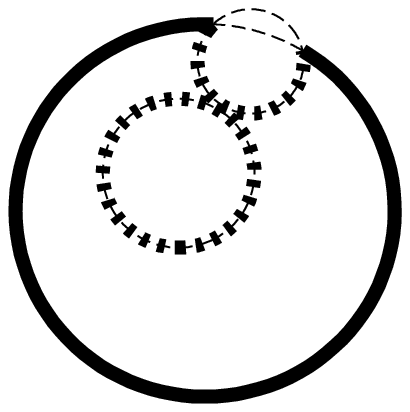}}%
\resizebox{20mm}{!}{\includegraphics{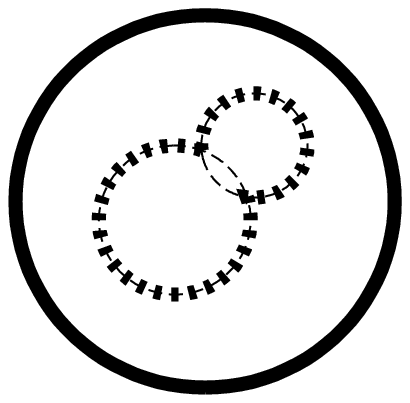}}%
\resizebox{20mm}{!}{\includegraphics{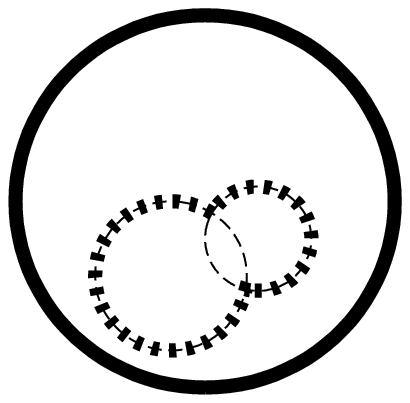}}%
\resizebox{20mm}{!}{\includegraphics{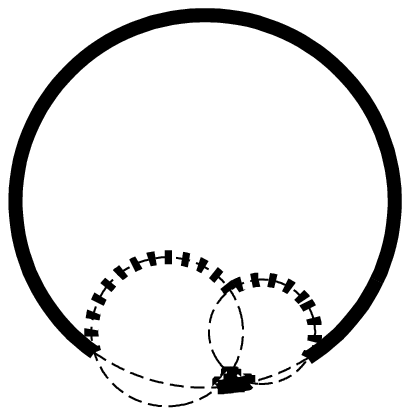}}%
\resizebox{20mm}{!}{\includegraphics{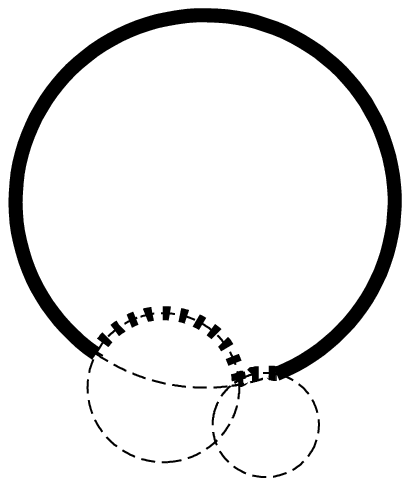}}%
\resizebox{20mm}{!}{\includegraphics{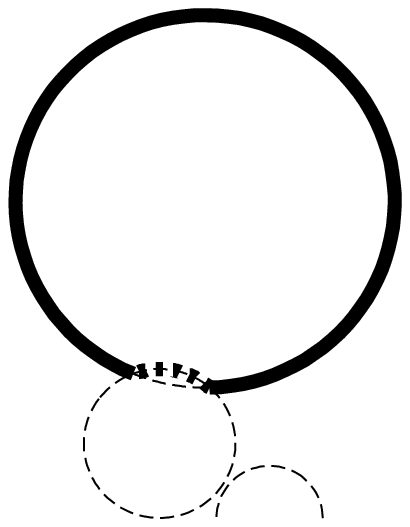}}
\end{center}
\caption{A ``movie'' of a mutual transit, caused by two relatively
large companion. The host star is the largest fixed circle while
the two companions are the smaller circles moving from up to down.
The arc boundaries of the area(s) on which the surface intensity is
integrated are denoted by thick lines. Thick solid lines mark where
the integrals in equation~(\ref{eq:intsum}) are directed 
in prograde (counter-clockwise) direction while thick dashed lines
mark where the integrals are directed in retrograde (clockwise)
direction. Thin lines mark the other arcs, which are irrelevant regarding
to the integration. Note that even in the topologically complex cases,
the number of relevant boundary arcs is three or less (with the exception
of the sixth frame where the number of relevant arcs is six).}
\label{fig:movie}\vspace*{-5mm} 
\end{figure*}

Let us consider a set arcs of which union is the boundary: $a \subset \partial S$.
The arc $a$ corresponding to the circle centered at $(x_a,y_a)$ with a radius
of $r_a$ and parameterized via the position angle $\varphi$ implies
the measure 
\begin{equation}
\mathrm{d}\mathbf{r}=\binom{-r\sin\varphi}{+r\cos\varphi}\mathrm{d}\varphi.
\end{equation}
Thus, the total flux $F$ coming from the area $S$ is then computed as
\begin{equation}
F=\sum\limits_{a\in\partial S}\int\limits_{\varphi^{(1)}_a}^{\varphi^{(2)}_a}
\left[f_y(x,y)\cos\varphi-f_x(x,y)\sin\varphi\right]r_a\mathrm{d}\varphi,\label{eq:intsum}
\end{equation}
where for more compact notations, we define $x\equiv x_a+r_a\cos\varphi$ and 
$y\equiv y_a+r_a\sin\varphi$. Note that the integration limits 
$\varphi^{(1)}_a$ and $\varphi^{(2)}_a$ are not necessarily
$\varphi^{(0)}_a$ and $\varphi^{(0)}_a+\Delta\varphi_a$, because
the direction of the $\oint\mathbf{f}\cdot\mathrm{d}\mathbf{r}$
integral must be positive in all cases. If we use 
$\varphi^{(1)}_a=\varphi^{(0)}_a$ and $\varphi^{(2)}_a=\varphi^{(0)}_a+\Delta\varphi_a$,
we have to multiply the integrand by $\pm 1$, depending whether the 
right-hand directed arc $a$ points inside or outside the area $S$
(see e.g. Fig.~\ref{fig:circles} or Fig.~\ref{fig:movie} for 
examples and further explanation).

\subsection{Some surface brightness functions}
\label{subsec:surfacebrightnessfunctions}

Now, we compute the integrals behind the sum of equation~(\ref{eq:intsum})
for various surface brightness functions. Let us consider a star
whose projected disk center is located at $(0,0)$ and has a radius
of unity. The domain of the functions of out interest is this 
unit circle, i.e. $x^2+y^2\le 1$. 

\subsubsection{Homogeneous surface}
\label{subsubsec:homogeneoussurface}

As we have seen earlier (see e.g. equation~\ref{eq:unitcurl}), 
the vector field $\mathbf{f}=\left(-\frac12y,+\frac12x\right)$ has a curl
of unity. For a given arc $a\equiv 0$, the integrals behind the 
sum of equation~(\ref{eq:intsum}) can be written as
\begin{eqnarray}
F_0 & = & \int\limits_{\varphi_1}^{\varphi_2}
\frac12(x_0+r\cos\varphi)r\cos\varphi+\frac12(y_0+r\sin\varphi)r\sin\varphi = \nonumber\\
& = & \int\limits_{\varphi_1}^{\varphi_2}
\frac12r(x_0\cos\varphi+y_0\sin\varphi)+\frac12r^2 = \nonumber\\
& = & \frac12r(\varphi_2-\varphi_2)+\frac12rx_0(\sin\varphi_2-\sin\varphi_1)+ \nonumber \\
& & +\frac12ry_0(\cos\varphi_1-\cos\varphi_2).\label{eq:inthomogeneous}
\end{eqnarray}
Note that the value of $F_0$ \emph{does} depend on the actual choice
for $\mathbf{f}$, i.e. it would be different if we add a gradient
to the vector field $\mathbf{f}$. However, $\sum\limits_a F_a$
in equation~(\ref{eq:intsum})
would not be altered after such an addition of a gradient field. By summing
the results yielded by equation~(\ref{eq:inthomogeneous}) for
the arcs $\{a\}$, we can easily reproduce the results of 
in Sections~2, 3 and Fig.~5 of \cite{kipping2011}.

\subsubsection{Polynomial intensities}
\label{subsubsec:polynomialintensities}

Various limb darkening models contain terms which can be quantified
as polynomial functions of the $(x,y)$ centroid coordinates (for instance,
the quadratic limb darkening law). In additional, any analytical limb
darkening profiles can be well approximated by polynomial functions, 
therefore it is worth to compute terms in equation~(\ref{eq:intsum})
for such cases. 

Without any restrictions, let us consider the term $x^py^q$. Due
to the linearity of the integral and summation, if the
surface intensity can be described by polynomials, computing
the integral in equation~(\ref{eq:intsum}) for the above terms
are sufficient. First, let us define
\begin{equation}
M_{pq} := (x_0+r\cos\varphi)^p(y_0+r\sin\varphi)^q,
\end{equation}
and introduce $c=\cos\varphi$ and $s=\sin\varphi$, just for simplicity.
Thus, in the expansion of equation~(\ref{eq:intsum}), we should compute
expressions like $\int M_{pq}c$ or $\int M_{pq}s$. 
Here we give a set of recurrence relations with which these indefinite
integrals can be evaluated. It is easy to show that
\begin{eqnarray}
\textstyle\int M_{pq} & = & x_0\textstyle\int M_{p-1,q}+r\textstyle\int M_{p-1,q}c, \hspace*{4mm} \text{or} \\
\textstyle\int M_{pq} & = & y_0\textstyle\int M_{p,q-1}+r\textstyle\int M_{p,q-1}s.
\end{eqnarray}
For the terms $\int M_{pq}c$ and $\int M_{pq}s$ we can write
\begin{eqnarray}
(1+p+q)\textstyle\int M_{pq} c & = & +M_{pq}s+rp\textstyle\int M_{p-1,q}+\\
	& & +x_0p\textstyle\int M_{p-1,q}c+y_0q\textstyle\int M_{p,q-1}c, \nonumber \\
(1+p+q)\textstyle\int M_{pq} s & = & -M_{pq}c+rq\textstyle\int M_{p,q-1}+\\
	& & +x_0p\textstyle\int M_{p-1,q}s+y_0q\textstyle\int M_{p,q-1}s. \nonumber
\end{eqnarray}
In order to bootstrap these set of recurrence relations,
we only have to use the following:
\begin{eqnarray}
M_{00} & = & 1, \\
\textstyle\int M_{00} & = & \textstyle\int 1 = \mathop{\mathrm{id}}, \\
\textstyle\int M_{00}c & = & +M_{00}s, \\
\textstyle\int M_{00}s & = & -M_{00}c.
\end{eqnarray}
However, for some cases we might compute these integrals a bit more
easier. For instance, the surface density $x^2+y^2$ can be integrated
as the exterior derivative of 
$\mathbf{f}=\left(-\frac12x^2y-\frac16y^3,+\frac16x^3+\frac12xy^2\right)$. Hence,
the primitive integral in equation~(\ref{eq:intsum}) for this $\mathbf{f}$
will be 
\begin{eqnarray}
F[\varphi] & = & \frac{r}{48}\left\{24(x_0^2+y_0^2)r+12r^3)\varphi\right. -  \\
	&   & -4y_0(6x_0^2+2y_0^2+9r^2)\cos\varphi + \nonumber\\
	&   & +4x_0(2x_0^2+6y_0^2+9r^2)\sin\varphi - \nonumber\\
	&   & -4r^2[y_0\cos(3\varphi)+x_0\sin(3\varphi)] - \nonumber\\
	&   & \left.-24x_0y_0r\cos(2\varphi)-r^3\sin(4\varphi)\right\}. \nonumber
\end{eqnarray}

\begin{figure*}
\begin{center}
\resizebox{80mm}{!}{\includegraphics{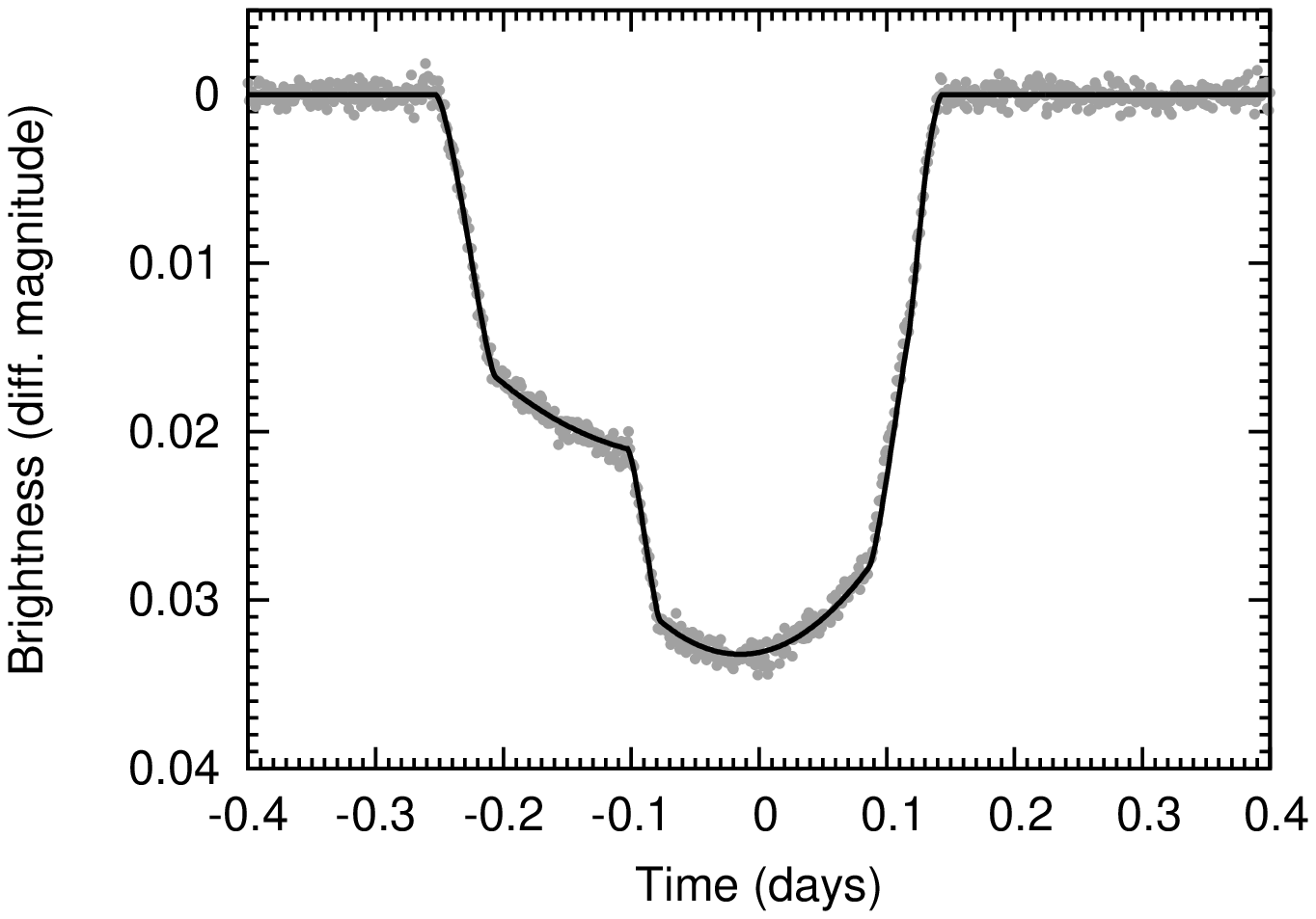}}%
\resizebox{80mm}{!}{\includegraphics{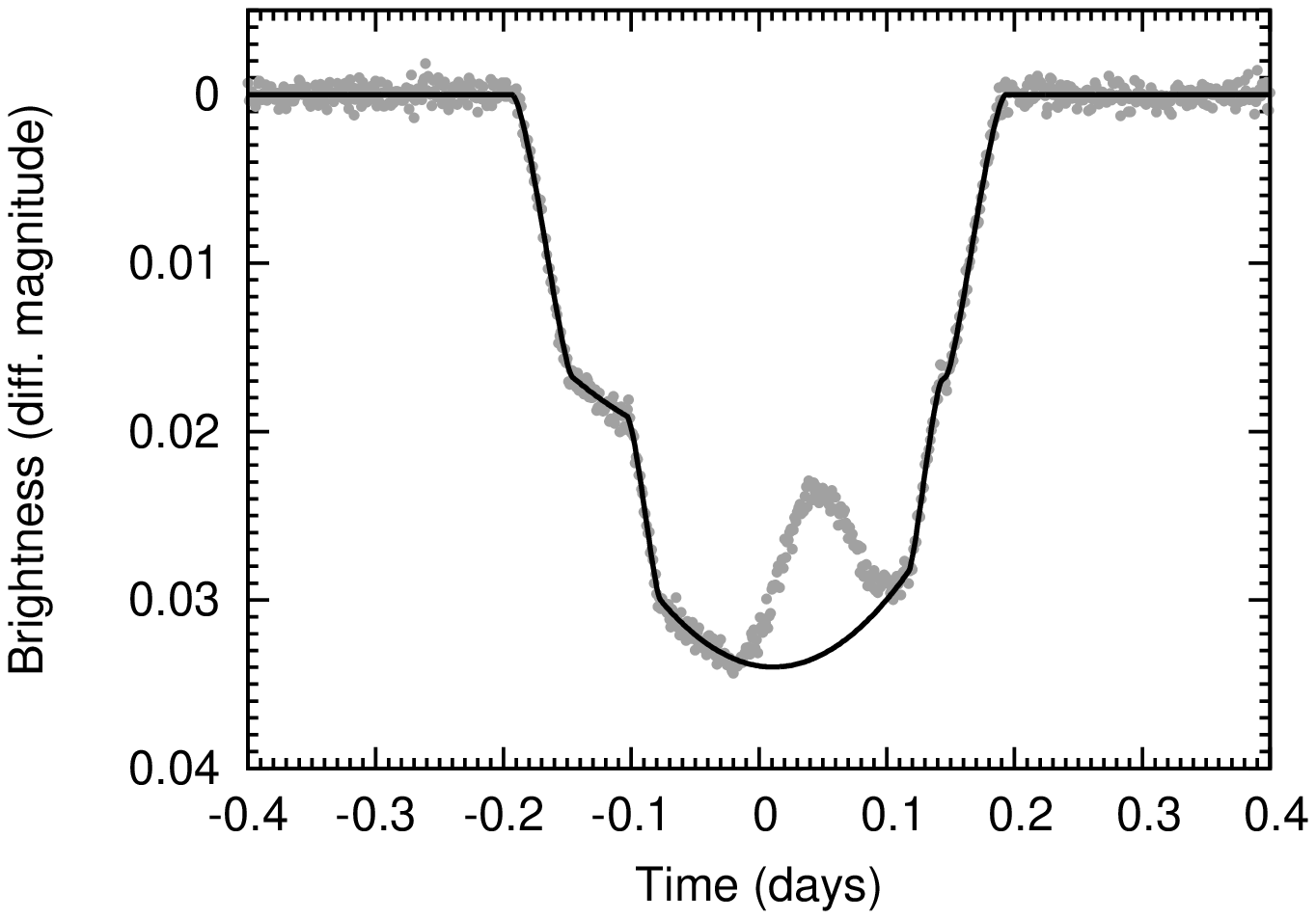}}
\end{center}
\caption{Simulated light curves of transits caused by two,
relatively large planets and occurring nearly simultaneously
but without any overlap (left panel). A light curve of a mutual 
transit with almost the same
geometry is displayed on the right panel. The signal-to-noise
ratio on these plots are nearly the same as one expects from 
Kepler photometry. The thick solid lines show the expected flux 
\emph{if} the two simultaneous transits would be treated independently,
i.e. the apparent overlapping of the planets would be neglected. 
See text for further details,
including the geometric parameters of these simulated transits.}
\label{fig:mutuallightcurve}
\end{figure*}

\subsubsection{Linear limb darkening}
\label{subsubsec:linearlimbdarkening}

The well known formula of linear limb darkening law gives us a surface
flux density that can be written in the form of 
\begin{equation}
I(x,y)=1-c[1-\mu],
\end{equation}
where $\mu=\sqrt{1-x^2-y^2}$ and $c$ denotes
the linear limb darkening coefficient. Performing integrals
on this function will yield a linear combination of integrating a 
constant (see Section~\ref{subsubsec:homogeneoussurface}) and integrating
the function $\sqrt{1-x^2-y^2}$. Thus, here we derive a function of which
exterior derivative is $\sqrt{1-x^2-y^2}$ as well as we compute the 
indefinite integral that is going to appear in equation~(\ref{eq:intsum}). 
First of all, we have to note that this function is defined only
in the domain bounded by the circle $x^2+y^2=1$. 

Let us search the function $\mathbf{f}$ of which exterior derivative 
is $\sqrt{1-x^2-y^2}\equiv\sqrt{1-r^2}$ in the form of 
$\mathbf{f}=\left[-yf(r^2)/2,+xf(r^2)/2\right]$. For simplicity, we
introduce $r^2=x^2+y^2$. 
It is easy to show that the expansion of equation~(\ref{eq:exterior}) for
this function yields an ordinary differential equation (ODE) for $f(\cdot)$
that is 
\begin{equation}
f(\xi)+\xi f'(\xi)=\sqrt{1-\xi},
\end{equation}
where $\xi=r^2$. One of the solutions of this ODE is
\begin{equation}
\frac{2}{3}\frac{1-(1-\xi)^{3/2}}{\xi}.
\end{equation}
This solution behaves analytically at $\xi=0$, i.e. at the center of the 
stellar disk. Therefore, the function $\mathbf{f}$ can be written as
\begin{equation}
\mathbf{f}=\binom{f_x}{f_y}=\frac{1-(1-x^2-y^2)^{3/2}}{3(x^2+y^2)}\binom{-y}{+x}.
\end{equation}
By substituting this $\mathbf{x}=(f_x,f_y)$ into equation~(\ref{eq:intsum}),
one finds that the primitive integral
\begin{equation}
F[\varphi']=\frac13\int\frac{1-(1-R^2-2r\rho\cos\varphi')^{3/2}}{R^2+2r\rho\cos\varphi'}(r+\rho\cos\varphi')r\,\mathrm{d}\varphi'\label{eq:llint}
\end{equation}
should be computed, if we parameterize the arc (on which 
the $\mathbf{f}$ is integrated) similarly as earlier.
The constants are the following:
$\rho^2=x_0^2+y_0^2$, $R^2=r^2+\rho^2$,  and 
$\varphi'$ is defined to be $\rho\cos\varphi'$ equals to $x_0\cos\varphi+y_0\sin\varphi$. 
The primitive integral in 
equation~(\ref{eq:llint}) can be computed analytically and this 
computation yields a function which contains elementary functions
as well as elliptic integrals. As a first step, let us introduce 
the constants 
\begin{eqnarray}
q_2 & = & r^2+\rho^2+2r\rho\cos\varphi', \label{eq:lpdef:first}\\
s_2 & = & (r+\rho)^2, \\
d_2 & = & (r-\rho)^2, \\
Q & = & \frac{1}{\sqrt{r\rho}}, \\
s_{\rm E} & = & 2\cos(\varphi'/2)\sqrt{\frac{r\rho}{1-d_2}}, \\
k_{\rm E} & = & \frac{1}{2}\sqrt{\frac{1-d_2}{r\rho}}, \\
n_{\rm E} & = & -\frac{1-d_2}{d_2}, \\
\hat F & = & \mathrm{F}(s_{\rm E};k_{\rm E}), \\
\hat E & = & \mathrm{E}(s_{\rm E};k_{\rm E}), \\
\hat P & = & \Pi(s_{\rm E};n_{\rm E},k_{\rm E}). \label{eq:lpdef:last}
\end{eqnarray}
Here $\mathrm{F}(\cdot;\cdot)$, $\mathrm{E}(\cdot;\cdot)$
and $\Pi(\cdot;\cdot,\cdot)$ denotes the incomplete elliptic integrals
of the first, second and third kind, respectively.
Then, $F[\varphi']$ in equation~(\ref{eq:llint}) is computed as 
\begin{eqnarray}
F[\varphi'] & = & 
	-\frac{1}{3}\arctan\left[\sqrt{\frac{d_2}{s_2}}\tan\left(\frac{\varphi'}{2}\right)\right]\frac{\rho^2-r^2}{\sqrt{d_2s_2}}+\nonumber \\
& + & \frac{\varphi'}{6}+\frac{2}{9}r\rho\sqrt{1-q_2}\sin\varphi'+ \nonumber \\
& + & \frac{1}{6}(1-4r^2+2r^4)Q\hat F+ \nonumber \\
& + & \frac{1}{9}r\rho(7r^2+5r\rho+\rho^2-4)Q\hat F+ \nonumber \\
& + & \frac{1}{9}r\rho(8-14r^2-2\rho^2)Q\hat E+ \nonumber \\
& + & \frac{1}{6}\frac{r+\rho}{r-\rho}Q\hat P.\label{eq:linprimitive}
\end{eqnarray}
One may note some similarities between these terms and the equations
in \cite{mandel2002} or \cite{pal2008}. Although the formulae above
include incomplete elliptic integrals, the actual evaluation of 
these does not require longer computation time
than the complete ones. Both types of elliptic integrals 
are computed via the Carlson symmetric forms \citep{carlson1993},
for which computation very fast and robust algorithms are 
available in the literature \citep{press1992,carlson1994}.

It should also be noted that the evaluation of equation~(\ref{eq:linprimitive})
might be done with caution in some cases where the values of 
the variables or constants defined in 
equations~(\ref{eq:lpdef:first})~-~(\ref{eq:lpdef:last})
introduce singularities in some of the terms. These values correspond
to cases where the arc endpoints are at the edge of the bounding circle
at $x^2+y^2=1$ and/or when the arc intersect the origin
(i.e. if $x=y=0$). However, these singularities
yield more simple formulae in general. For instance, the case of $\rho=0$
(i.e. when the bounding circle and the arc is concentric), 
equation~(\ref{eq:linprimitive}) becomes simply
\begin{equation}
F[\varphi']=\frac13\left[1-(1-r^2)^{3/2}\right]\varphi'.
\end{equation}
Especially, when the arc is a part of the bounding circle itself
(which is a practically frequent case: see e.g. the thick solid 
lines in the frames of Fig.~\ref{fig:movie}), 
i.e. when $r=1$ and $\rho=0$, then $F[\varphi']$ is merely $\varphi'/3$.
All in all, these cases should be treated 
carefully during a practical implementation. 

\subsubsection{Quadratic limb darkening}
\label{subsubsec:quadraticlimbdarkening}

The quadratic limb darkening stellar profile is characterized by 
the surface flux density $I=1-c_1(1-\mu)-c_2(1-\mu)^2$.
Since $\mu=\sqrt{1-x^2-y^2}$, by expanding this equation, 
we obtain a constant term, with a
value of $1-c_1-2c_2$, a polynomial term $x^2+y^2$ with a coefficient
$+c_2$ and a term that is proportional to $\mu$
and has a coefficient $c_1+2c_2$. Hence, the formulae in the previous
three subsections (\ref{subsubsec:homogeneoussurface},
\ref{subsubsec:polynomialintensities} and \ref{subsubsec:linearlimbdarkening})
can be applied accordingly to evaluate the final apparent
fluxes in the case of a quadratic limb darkening law. 


\section{Orbital inclinations}
\label{sec:mutualtransits}

Mutual transits occur when at least two bodies (that can be, for instance,
two planets or a planet and its moon) transits the host star simultaneously
\emph{and} their projections also overlap. Due to this overlap,
the observed flux coming from the host star is \emph{larger} than 
if we would consider naively the flux decreases from each body
independently. In Fig.~\ref{fig:movie} we display a series of
images that clearly show this effect. In the previous section we
deduced the algorithms and mathematical formulae that are
needed for the computation of the total observed flux for
arbitrary geometry and for various limb darkening models.

As a demonstration, in Fig.~\ref{fig:mutuallightcurve} 
we display two simulated 
light curves with nearly the same orbital geometry. The planet-to-size
ratio for the two companions are $R_1/R_\star=0.13$
and $R_2/R_\star=0.10$ while the orbital parameters
are the following: $a_1/R_\star=4.3000$, $b_1=0.35$, $n_1=2.0\,{\rm d}^{-1}$,
$a_2/R_\star=9.5952$, $b_2=0.22$, $n_2=0.6\,{\rm d}^{-1}$, 
$\Delta\Omega=18^\circ$ and both of the planets have a circular orbit.
Here $n_k$ denotes the orbital angular frequency: it is $n_k=2\pi/P_k$,
where $P_k$ is the orbital period. $b_k$ is the impact parameter
of the transit, $a_k/R_\star$ is the normalized semimajor axis 
(in the units of stellar radii) and $\Delta\Omega=\Omega_1-\Omega_2$ 
is difference in the orbital ascending nodes (note that the reference
plane here is the plane of the sky). 
The mid-transit time of the inner planet is $E_1=0.02\,{\rm d}$ while
the outer planet has $E_2=-0.06\,{\rm d}$ on the left panel, and
$E_2=0.00\,{\rm d}$ on the right panel.
This difference between the mid-transit times yields a mutual transit 
in the latter case (see the flux excess in Fig.~\ref{fig:mutuallightcurve}
at $t\approx0.02\dots0.08\,{\rm d}$)
while there is no overlap between the apparent planetary
disks in the former case. 

It can easily be seen that the time evolution of the flux excess
yielded by the mutual transit\footnote{Here we treat this ``flux excess''
relative to the flux level that would be if we neglect the effect
of the overlapping and simply calculate the yield of the two components
independently.} has similar qualitative 
properties as the normal transits have. Namely, it has a mid-time, a
peak and a duration. The larger the flux excess peak, the larger 
the overlapping area is. At a first glance, the only quantity
for which an observation of a mutual transit yields additional
constraints is the difference in the $\Delta\Omega$, the 
difference between the orbital ascending nodes. Qualitatively,
the longer the duration of this flux excess, the smaller 
the absolute value of $\Delta\Omega$ is\footnote{Imagine two completely
retrograde orbits: in this case, the relative speed of the 
transiting planets is the highest, thus the duration of the overlapping
will be the smallest.}. However, the depth and the exact time of 
the mutual event defines the impact parameters more precisely.
This is rather relevant when one or both of the impact parameters
are relatively small: the uncertainty of $b^2$ does not 
strongly depend on the actual value of $b$ 
\citep[see e.g.][]{pal2008,carter2008}, thus the uncertainty 
in $b$ will be rather large for small values of $b$ due
to the relation $\Delta b=(2b)^{-1}\Delta(b^2)$. Indeed, for instance,
the analysis of the light curves shown in Fig.~\ref{fig:mutuallightcurve}
yields the following. If no mutual transit occurs (left panel), the best-fit
values for $b_k$'s will be $b_1=0.323\pm0.012$ and $b_2=0.215\pm0.016$
while if we can observe the mutual transit, we obtain 
$b_1=0.351\pm0.003$ and $b_2=0.223\pm0.007$ while for the
node difference we got $\Delta\Omega=17.4\pm0.5^\circ$. 
For this demonstration of light curve analysis, we employed an
improved Markov Chain Monte-Carlo algorithm as implemented
in the \texttt{lfit} utility \citep{pal2009}.

Of course, if the difference in the nodes, $\Delta\Omega=\Omega_1-\Omega_2$
is known, we can compute the mutual inclination $i_{\rm m}$ 
of the orbits as well using the well-known relation
\begin{equation}
\cos i_{\rm m}=\cos i_1\cos i_2+\sin i_1\sin i_2\cos\Delta\Omega.
\end{equation}
It should also be mentioned that the analysis of mutual transits resolve
the ambiguity between the values of $\pm\Delta\Omega$. 
And of course, the precise analysis of mutual transits should involve
the gravitational interactions between the companions
\citep[see e.g.][]{pal2010}, especially when data are available on 
a timescale on which the perturbations are not negligible (contrary
to the demonstration presented here).


\section{Discussion}
\label{sec:summary}

In this paper we investigated the possibilities for computing 
apparent stellar fluxes in multiple or hierarchical stellar and/or
planetary systems during simultaneous transits or occultations. 
The presented algorithm is capable to derive these fluxes for 
arbitrary number of bodies that are actively parts of the
transiting or eclipsing event. This method can then be applied
for various analyses of complex astrophysical systems, including
multiple transiting planetary systems, hierarchical stellar systems
with planetary companions and extrasolar moons as well. 

Currently, the algorithm is implemented in ANSI C, in the form
of a plug-in module for the program \texttt{lfit} and available 
from the address \texttt{http://szofi.elte.hu/\~{ }apal/utils/astro/mttr/}.
This module features functions named \texttt{mttrXy(.)}, where 
\texttt{X} denotes the number of transiting bodies and \texttt{y}
can be ``\texttt{u}'', ``\texttt{l}'' or ``\texttt{q}'' for
the uniform flux density, linear limb darkening and quadratic limb
darkening. Evidently, these functions have $3X+y$ parameters where
$y$ is $0$, $1$ or $2$ for ``\texttt{u}'', ``\texttt{l}'' or ``\texttt{q}'',
respectively. 
The current version of this module does not compute the parametric
derivatives of the functions analytically but emulates them using 
numerical approximations for the \texttt{lfit} utility. Since both
the parametric derivatives of the arcs (with respect to the 
circle center coordinates and radii) and the parametric derivatives
of equation~(\ref{eq:intsum}) can be computed analytically, the composition
of these two would give us the required derivatives. 

As a demonstration, we applied this method to obtain mutual inclinations
of orbits in multiple transiting planetary systems. The analysis presented
here clearly shows that observing a mutual transits yields not only 
an accurate value for the ascending node difference but also results
a more precise value for the impact parameters, and therefore the 
orbital inclinations as well.


\section*{Acknowledgments}

The author would like to thank the anonymous referee for the valuable 
suggestions and comments. The work of the author has been supported 
by the ESA grant PECS~98073 and by the J\'anos Bolyai Research 
Scholarship of the Hungarian Academy of Sciences. 


{}

\bsp


\label{lastpage}

\end{document}